\documentclass[11pt]{article}%
\usepackage{amsmath}
\usepackage{graphicx}%
\usepackage{amsfonts}%
\usepackage{amssymb}
\newcommand{\eqnb}{\begin{equation}}
\newcommand{\eqne}{\end{equation}}

\newtheorem{The}{Theorem}

\newtheorem{Def}{Definition}

\newtheorem{Rem}{Remark}

\begin{document}

\title{\textbf{Nonlinear Markov Processes in Big Networks}}
\author{Quan-Lin Li\\School of Economics and Management Sciences \\Yanshan University, Qinhuangdao 066004, P.R. China}
\date{Published in {\bf Special Matrices}, 2016, Vol. 4, 202--217}
\maketitle

\begin{abstract}
Big networks express multiple classes of large-scale networks in many
practical areas such as computer networks, internet of things, cloud
computation, manufacturing systems, transportation networks, and healthcare
systems. This paper analyzes such big networks, and applies the mean-field
theory and the nonlinear Markov processes to constructing a broad class of
nonlinear continuous-time block-structured Markov processes, which can be used
to deal with many practical stochastic systems. Firstly, a nonlinear Markov
process is derived from a large number of big networks with weak interactions,
where each big network is described as a continuous-time block-structured
Markov process. Secondly, some effective algorithms are given for computing
the fixed points of the nonlinear Markov process by means of the UL-type
$RG$-factorization. Finally, the Birkhoff center, the locally stable fixed
points, the Lyapunov functions and the relative entropy are developed to
analyze stability or metastability of the system of weakly interacting big
networks, and several interesting open problems are proposed with detailed
interpretation. We believe that the methodology and results given in this
paper can be useful and effective in the study of big networks.

\vskip                                                            0.5cm


\end{abstract}

\section{Introduction}

In this paper, we consider a large number of big networks with weak
interactions, where each big network is described as a continuous-time
block-structured Markov process, which can be applied to deal with many
practical stochastic systems. As the number of big networks goes to infinite,
the interactions between any two subsets of the big networks become negligible
or are asymptotically independent, and the overall effect of the interactions
can be replaced by an empirical measure under a mean-field setting. Based on
this, the evolution of such a big network is expressed as a time-inhomogeneous
block-structured Markov process, which leads to that transient performance of
the big network can be discussed by means of a system of ordinary differential
equations, while its stationary performance measures can be computed in terms
of a fixed point, which satisfies a system of nonlinear equations.

The purpose of this paper is to develop the mean-field computational theory
both for performance evaluation and for performance optimization of big
networks. During the last three decades considerable attention has been paid
to studying the mean-field theory of large-scale stochastic systems, which has
been well documented, for example, an excellent survey paper by Sznitman
\cite{Szn:1989}; interacting Markov processes by, such as, Spitzer
\cite{Spi:1970}, Kurtz \cite{Kur:1970, Kur:1971, Kur:1978}, Dawson
\cite{Daw:1983}, Shiga and Tanaka \cite{Shi:1985}, Dawson and Zheng
\cite{Daw:1991}, Duffield and Werner \cite{Duf:1991}, Duffield \cite{Duf:1992}%
, Le Boudecet at al. \cite{Le:2007}, Darling and Norris
\cite{Dar:2008}, Bordenave at al. \cite{Bor:2010}, Benaim and Le
Boudec \cite{Ben:2008, Ben:2011} and Li \cite{Li:2014}; interacting
Markov decision processes by Gast and Gaujal \cite{Gas:2011} and
Gast at al. \cite{Gas:2012}; and more generally, interacting
particle systems introduced by three books of Kipnis and Landim
\cite{Kip:1999}, Chen \cite{Chen:2004} and Liggett \cite{Lig:2005}.
For limit theory of stochastic process sequences, readers may refer
to three books of Ethier and Kurtz \cite{Eth:1986}, Whitt
\cite{Whi:2002}, and Jacod and Shiryaev \cite{Jac:2003}.

During the last two decades the mean-field theory has widely been applied to
studying some practical networks (or systems) including queueing systems,
communication networks, manufacturing systems, transportation networks and so
forth. (a) For queueing systems, Baccelli et al. \cite{Bac:1992} and Kelly
\cite{Kel:1991} first applied the mean-field theory to the study of queueing
networks. Subsequent papers have been published on this theme, among which see
Borovkov \cite{Bor:1998}, Bobbio et al. \cite{Bob:2008}; Delcoigne and Fayolle
\cite{Del:1999} for polling systems; Karpelevich and Rybko \cite{Kar:2000} for
symmetric closed queuing networks; Baccelli et al. \cite{Bac:2013} for varying
topology networks; Hayden et al. \cite{Hay:2012} for stochastic process
algebra; and Hunt and Kurtz \cite{Hun:1994}, and Zachary and Ziedins
\cite{Zac:2002} for large loss networks. (b) To discuss randomized load
balancing, some work has been done on two different research directions:
Supermarket models, and work stealing models. For the supermarket models,
readers may refer to, such as, Vvedenskaya et al. \cite{Vve:1996} for operator
semigroup, Mitzenmacher \cite{Mit:1996} for density-dependent jump Markov
processes, Turner \cite{Tur:1996} for martingale limits. Subsequent papers
have been published on this theme, important examples include Vvedenskaya and
Suhov \cite{Vve:1997}, Graham \cite{Gra:2000, Gra:2004}, Luczak and McDiarmid
\cite{Luc:2006, Luc:2007}, Li et al. \cite{LiD:2014, LiD:2015} and Li
\cite{LiL:2014}. As a key generalization of the supermarket models, the fast
Jackson networks were investigated by Martin and Suhov \cite{Mar:1999}, Martin
\cite{Mar:2001}, and Suhov and Vvedenskaya \cite{Suh:2002}. In contrast, the
available results of the work stealing models based on the mean-field theory,
Markov processes and queueing theory are still few in the literature, e.g.,
see Gast and Gaujal \cite{Gas:2010} and Li and Yang \cite{LiY:2015}. In
addition, readers may refer to the computer and communication systems by
Benaim and Le Boudec \cite{Ben:2008, Ben:2011}, and Antunes et al.
\cite{Ant:2006, Ant:2008}; the bike sharing systems by Fricker et al.
\cite{Fri:2012} and Fricker and Gast \cite{Fri:2014}; and the transportation
networks by Oseledets and Khmelev \cite{Ose:2002}.

Nonlinear Markov processes play an important role in the study of big
networks. Important examples include Rybko and Shlosman \cite{Ryb:2003}, Peng
\cite{Peng:2005}, Turner \cite{Tur:2007}, Benaim and Le Boudec \cite{Ben:2008}%
, Frank \cite{Fra:2008}, Kolokoltsov \cite{Kol:2010}, Gast and Gaujal
\cite{Gas:2011}, Kolokoltsov \cite{Kol:2010, Kol:2011a}, Kolokoltsov at al.
\cite{Kol:2011}, Muzychka and Vaninsky \cite{Muz:2011}, Dupuis and Fischer
\cite{Dup:2011}, Gast at al. \cite{Gas:2012}, Vaninsky et al. \cite{Van:2012},
Budhiraja et al. \cite{Bud:2014, Bud:2014a}, Budhiraja and Majumder
\cite{BudM:2014} and Benaim \cite{Ben:2014}.

Metastability is an ubiquitous and important phenomenon of the dynamical
behavior of communication networks, e.g., see Marbukh \cite{Mar:1984}, Gibbens
at al. \cite{Gib:1990}, Kelly \cite{Kel:1991} (see Page 349), Dobrushin
\cite{Dob:1993}, Antunes et al. \cite{Ant:2008} and Tibi \cite{Tib:2010}. For
metastability in Markov processes, readers may refer to Galves at al.
\cite{Gal:1987}, Bovier et al. \cite{Bov:2001, Bov:2002}, Olivieri and Vares
\cite{Oli:2005}, Freidlin and Wentzell \cite{Fre:1984}, Bovier \cite{Bov:2003,
Bov:2006}, den Hollander \cite{den:2004}, and Beltran and Landim
\cite{Bel:2010, Bel:2012}.

The main contributions of this paper are threefold. The first one is to set up
a broad class of nonlinear continuous-time block-structured Markov processes
when applying the mean-field theory to analysis of big networks with weak
interactions. The second one is to propose some effective algorithms for
computing the fixed points of the nonlinear Markov processes by means of the
UL-type $RG$-factorization, and show for some big networks that there possibly
exist multiple fixed points, which lead to the metastability. The third one is
to use the Birkhoff center, the locally stable fixed points, the Lyapunov
functions and the relative entropy to analyze either stability or
metastability of the big networks, and to give several interesting open
problems with detailed interpretation. Furthermore, this paper provides a new
method for computing the locally stable fixed points in the study of big
networks. We believe that the methodology and results given in this paper can
be useful and effective in performance evaluation and performance optimization
of big networks.

The remainder of this paper is organized as follows. In Section 2, we derive a
class of nonlinear Markov processes through an asymptotic analysis of the
weakly interacting big networks. In Section 3, we provide some effective
algorithms for computing the fixed points of the dynamic system of mean-field
equations. In Section 4, we discuss the Birkhoff center and the locally stable
fixed points of the dynamic system of mean-field equations, and apply the
Lyapunov functions and the relative entropy to study the stability or
metastability of the big network. Also, we provide several interesting open
problems with detailed interpretation. Some concluding remarks are given in
the final section.

\section{Nonlinear Markov Processes}

In this section, we derive a class of nonlinear Markov processes through an
asymptotic analysis for a collection of weakly interacting big networks, in
which each big network evolves as a continuous-time block-structured Markov
process, which can be applied to deal with many practical stochastic systems.

To discuss a system of weakly interacting big networks, we assume that any
individual of the big networks evolves as a continuous-time block-structured
Markov process $\mathcal{X}$ whose infinitesimal generator is given by%
\begin{equation}
Q=\left(
\begin{array}
[c]{ccccc}%
Q_{0,0} & Q_{0,1} & Q_{0,2} & Q_{0,3} & \cdots\\
Q_{1,0} & Q_{1,1} & Q_{1,2} & Q_{1,3} & \cdots\\
Q_{2,0} & Q_{2,1} & Q_{2,2} & Q_{2,3} & \cdots\\
Q_{3,0} & Q_{3,1} & Q_{3,2} & Q_{3,3} & \cdots\\
\vdots & \vdots & \vdots & \vdots &
\end{array}
\right)  , \label{gen-1}%
\end{equation}
where the size of the matrix $Q_{j,j}$ is $m_{j}$ for $j\geq0$, and the sizes
of other matrices can be determined accordingly. It is easy to see that the
matrix $Q_{j,j}$ is also the infinitesimal generator of a continuous-time
Markov process with $m_{j}$ states for $j\geq0$. We assume that the
continuous-time Markov process $Q$ is irreducible, aperiodic and positive
recurrent, and its state space may be expressed as a two-dimensional
structure: $\Omega=\left\{  \left(  k,j\right)  :k\geq0,1\leq j\leq
m_{k}\right\}  $. See Li \cite{Li:2010} for more details.

From the continuous-time block-structured Markov chain $\mathcal{X}$, the
system of $N$ weakly interacting big networks is described as an
$\mathcal{X}^{N}$-valued Markov process, and the states of the $N$ big
networks are denoted as $X^{1,N}\left(  t\right)  $, $X^{2,N}\left(  t\right)
$, $\ldots$, $X^{N,N}\left(  t\right)  $, respectively.

Let $\mathbf{X}^{N}\left(  t\right)  =\left(  X^{1,N}\left(  t\right)
,X^{2,N}\left(  t\right)  ,\ldots,X^{N,N}\left(  t\right)  \right)  $. Then
the empirical measure of the system of $N$ weakly interacting big networks is
given by%
\begin{equation}
\mu^{N}\left(  t\right)  =\frac{1}{N}\sum_{i=1}^{N}\delta_{X^{i,N}\left(
t\right)  }, \label{emp}%
\end{equation}
where $\delta_{x}$ is the Dirac measure at $x$.

We denote by $\mathfrak{P}\left(  \Omega\right)  $ the space of probability
vectors on the state space $\Omega$, which is equipped with the usual topology
of weak convergence. If $p\in\mathfrak{P}\left(  \Omega\right)  $, we write
$p=\left(  p_{0},p_{1},p_{2},\ldots\right)  $, where the size of the vector
$p_{j}$ is $m_{j}$ for $j\geq0$. At the same time, it is clear that $\mu
^{N}\left(  t\right)  \in\mathfrak{P}\left(  \Omega\right)  $ is a random
variable for $t\geq0$, and $\left\{  \mu^{N}\left(  t\right)  :t\geq0\right\}
$ is a continuous-time Markov process.

For the $\mathcal{X}^{N}$-valued continuous-time block-structured Markov
process, we define that the probability distribution of $\mathbf{X}^{N}\left(
t\right)  $ is exchangeable, if for any level permutation $(k_{i_{1}}%
,k_{i_{2}},\ldots$, $k_{i_{N}})$ of $\left(  k_{1},k_{2},\ldots,k_{N}\right)
$ and any phase permutation $\left(  j_{i_{1}},j_{i_{2}},\ldots,j_{i_{N}%
}\right)  $ of $\left(  j_{1},j_{2},\ldots,j_{N}\right)  $,
\begin{align}
&  P\left\{  X^{1,N}\left(  t\right)  =\left(  k_{1},j_{1}\right)
,X^{2,N}\left(  t\right)  =\left(  k_{2},j_{2}\right)  ,\ldots,X^{N,N}\left(
t\right)  =\left(  k_{N},j_{N}\right)  \right\} \label{exch}\\
&  =P\left\{  X^{i_{1},N}\left(  t\right)  =\left(  k_{i_{1}},j_{i_{1}%
}\right)  ,X^{i_{2},N}\left(  t\right)  =\left(  k_{i_{2}},j_{i_{2}}\right)
,\ldots,X^{i_{N},N}\left(  t\right)  =\left(  k_{i_{N}},j_{i_{N}}\right)
\right\}  .\nonumber
\end{align}

In the system of $N$ weakly interacting big networks, the effect of a tagged
big network on the dynamics of the system of $N$ weakly interacting big
networks is of order $1/N$, and the jump intensity of any given big network
depends on the configuration of other big networks only through the empirical
measure $\mu^{N}\left(  t\right)  $. To study the system of $N$ weakly
interacting big networks, it is seen from probability one that at most one big
network will jump, i.e., change state, at a given time, and the jump
intensities of any given big network depend only on its own state and the
state of the empirical measure at that time. In addition, the jump intensities
of the $N$ weakly interacting big networks have the same functional form.
Based on this, for the $\mathcal{X}^{N}$-valued Markov process, if the initial
probability distribution of $\mathbf{X}^{N}\left(  0\right)  $ is
exchangeable, then at any time $t\geq0$, the probability distribution of
$\mathbf{X}^{N}\left(  t\right)  $ is also exchangeable.

For the system of $N$ weakly interacting big networks, if the probability
distribution of $\mathbf{X}^{N}\left(  t\right)  $ is exchangeable, then the
$N$ big networks are indistinguishable, thus we apply the mean-field theory to
discussion of this system through only considering the Markov process of a
tagged big network (such as, the first big network); while analysis of the
total system can be completed by the propagation of chaos (as $N\rightarrow
\infty$). Based on this, the infinitesimal generator of the Markov process
corresponding to the tagged big network is defined as%
\begin{equation}
\Gamma^{\left(  N\right)  }\left(  \mu^{N}\left(  t\right)  \right)  =\left(
\begin{array}
[c]{ccccc}%
\Gamma_{0,0}^{\left(  N\right)  }\left(  \mu^{N}\left(  t\right)  \right)  &
\Gamma_{0,1}^{\left(  N\right)  }\left(  \mu^{N}\left(  t\right)  \right)  &
\Gamma_{0,2}^{\left(  N\right)  }\left(  \mu^{N}\left(  t\right)  \right)  &
\Gamma_{0,3}^{\left(  N\right)  }\left(  \mu^{N}\left(  t\right)  \right)  &
\cdots\\
\Gamma_{1,0}^{\left(  N\right)  }\left(  \mu^{N}\left(  t\right)  \right)  &
\Gamma_{1,1}^{\left(  N\right)  }\left(  \mu^{N}\left(  t\right)  \right)  &
\Gamma_{1,2}^{\left(  N\right)  }\left(  \mu^{N}\left(  t\right)  \right)  &
\Gamma_{1,3}^{\left(  N\right)  }\left(  \mu^{N}\left(  t\right)  \right)  &
\cdots\\
\Gamma_{2,0}^{\left(  N\right)  }\left(  \mu^{N}\left(  t\right)  \right)  &
\Gamma_{2,1}^{\left(  N\right)  }\left(  \mu^{N}\left(  t\right)  \right)  &
\Gamma_{2,2}^{\left(  N\right)  }\left(  \mu^{N}\left(  t\right)  \right)  &
\Gamma_{2,3}^{\left(  N\right)  }\left(  \mu^{N}\left(  t\right)  \right)  &
\cdots\\
\Gamma_{3,0}^{\left(  N\right)  }\left(  \mu^{N}\left(  t\right)  \right)  &
\Gamma_{3,1}^{\left(  N\right)  }\left(  \mu^{N}\left(  t\right)  \right)  &
\Gamma_{3,2}^{\left(  N\right)  }\left(  \mu^{N}\left(  t\right)  \right)  &
\Gamma_{3,3}^{\left(  N\right)  }\left(  \mu^{N}\left(  t\right)  \right)  &
\cdots\\
\vdots & \vdots & \vdots & \vdots &
\end{array}
\right)  , \label{gen-2}%
\end{equation}
where the size of the matrix $\Gamma_{j,j}^{\left(  N\right)  }\left(  \mu
^{N}\left(  t\right)  \right)  $ is $m_{j}$ for $j\geq0$, and the sizes of
other matrices can be determined accordingly. Since $\mu^{N}\left(  t\right)
$ is a random variable, it is clear that $\Gamma^{\left(  N\right)  }\left(
\mu^{N}\left(  t\right)  \right)  $ is a random matrix of infinite order. On
the other hand, it is seen from the law of large number that the limit of the
empirical measure $\mu^{N}\left(  t\right)  $ is deterministic under suitable conditions.

Now, we analyze some convergence of the sequence $\left\{
\mu^{N}\left( t\right)  :t\geq0\right\}  $ of Markov processes for
$N=1,2,3,\ldots$, and our aim is to provide a basic support for our
later study of various convergence involved. To this end, we
consider the empirical measure: $\mu^{N}\left( t\right)
=\frac{1}{N}\sum_{i=1}^{N}\delta_{X^{i,N}\left(  t\right)  }$ with
samples in $\mathcal{P}\left(  \mathbb{D}\left(
R_{+},\mathbf{N}\right) \right)  $, where $R_{+}=[0,+\infty)$,
$\mathbf{N=}\left\{  \left( k,j\right)  :k\geq0,1\leq j\leq
m_{k}\right\}  $, and $\mathbb{D}\left( R_{+},\mathbf{N}\right)  $
is the Skorohod space, i.e., the set of mappings
$x:R_{+}\rightarrow\mathbf{N}$ which are right continuous with
left-hand limits (in short, c\`{a}dl\`{a}g). Readers may refer to
Chapter 3 of Ethier and Kurtz \cite{Eth:1986} for more details.
Notice that the convergence in the Skorohod topology means the
convergence in distribution (or weak convergence) for the Skorohod
topology on the space of trajectories, we assume that the sequence
$\left\{  \mu^{N}\left(  t\right)  :t\geq0\right\}  $ of Markov
processes converges in probability (or converges weakly), for the
Skorohod topology, to a given probability vector $p\left(  t\right)
$. At the same time, for this weak convergence, we write
$\mu^{N}\left(  t\right) \Longrightarrow p\left(  t\right)  $ for
$t\geq0$, as $N\rightarrow\infty$.

Let $\mu^{N}\left(  t\right)  \Longrightarrow p\left(  t\right)  $ and
$\Gamma^{\left(  N\right)  }\left(  \mu^{N}\left(  t\right)  \right)
\Longrightarrow\Gamma\left(  p\left(  t\right)  \right)  $ for $t\geq0$, as
$N\rightarrow\infty$. Then $p\left(  t\right)  $ is a given probability
vector. Furthermore, using some probability analysis, we may obtain an
infinite-dimensional dynamic system of mean-field equations as follows:
\begin{equation}
\frac{\text{d}}{\text{d}t}p\left(  t\right)  =p\left(  t\right)  \Gamma\left(
p\left(  t\right)  \right)  \label{dyn-1}%
\end{equation}
with the initial condition%
\begin{equation}
p\left(  0\right)  =q. \label{dyn-2}%
\end{equation}
Obviously, the dynamic system of mean-field equations, given in (\ref{dyn-1})
and (\ref{dyn-2}), is related to a nonlinear Markov process whose
infinitesimal generator is given by%
\begin{equation}
\Gamma\left(  p\left(  t\right)  \right)  =\left(
\begin{array}
[c]{ccccc}%
\Gamma_{0,0}\left(  p\left(  t\right)  \right)  & \Gamma_{0,1}\left(  p\left(
t\right)  \right)  & \Gamma_{0,2}\left(  p\left(  t\right)  \right)  &
\Gamma_{0,3}\left(  p\left(  t\right)  \right)  & \cdots\\
\Gamma_{1,0}\left(  p\left(  t\right)  \right)  & \Gamma_{1,1}\left(  p\left(
t\right)  \right)  & \Gamma_{1,2}\left(  p\left(  t\right)  \right)  &
\Gamma_{1,3}\left(  p\left(  t\right)  \right)  & \cdots\\
\Gamma_{2,0}\left(  p\left(  t\right)  \right)  & \Gamma_{2,1}\left(  p\left(
t\right)  \right)  & \Gamma_{2,2}\left(  p\left(  t\right)  \right)  &
\Gamma_{2,3}\left(  p\left(  t\right)  \right)  & \cdots\\
\Gamma_{3,0}\left(  p\left(  t\right)  \right)  & \Gamma_{3,1}\left(  p\left(
t\right)  \right)  & \Gamma_{3,2}\left(  p\left(  t\right)  \right)  &
\Gamma_{3,3}\left(  p\left(  t\right)  \right)  & \cdots\\
\vdots & \vdots & \vdots & \vdots &
\end{array}
\right)  . \label{gen-3}%
\end{equation}

\begin{Rem}
To establish the infinitesimal generator $\Gamma\left(  p\left(  t\right)
\right)  $ of a nonlinear Markov process, readers may also refer to some
recent publications, for example, the discrete-time Markov chains by Benaim
and Le Boudec \cite{Ben:2008} and Budhiraja and Majumder \cite{BudM:2014}, the
Markov decision processes by Gast and Gaujal \cite{Gas:2011} and Gast at al.
\cite{Gas:2012}, the continuous-time Markov chains by Dupuis and Fischer
\cite{Dup:2011} and Budhiraja et al. \cite{Bud:2014, Bud:2014a}, and some
practical examples include Mitzenmacher \cite{Mit:1996}, Bobbio et al.
\cite{Bob:2008}, Li et al. \cite{LiD:2014, LiD:2015}, and Li and Lui
\cite{LiL:2014}.
\end{Rem}

In what follows, it is necessary to provide some useful interpretation or
proofs for how to establish the dynamic system of mean-field equations
(\ref{dyn-1}) and (\ref{dyn-2}).

\textbf{(a) Existence and Uniqueness}

Consider the infinite-dimensional ordinary differential equation:
$\frac{\text{d}}{\text{d}t}p\left(  t\right)  =p\left(  t\right)
\Gamma\left(  p\left(  t\right)  \right)  $ with $p\left(  0\right)  =q$. A
solution in the classical sense is a (continuously) differential function
$p\left(  t\right)  $ such that $\frac{\text{d}}{\text{d}t}p\left(  t\right)
=p\left(  t\right)  \Gamma\left(  p\left(  t\right)  \right)  $ with $p\left(
0\right)  =q$. A classical method is the Picard approximation as follows. If
$\Gamma\left(  x\right)  $ is (locally) Lipschitz on a set $E\subseteq
\mathfrak{P}\left(  \Omega\right)  $, that is, there exists a positive
constant $C$ such that%
\[
\left\|  \Gamma\left(  x\right)  -\Gamma\left(  y\right)  \right\|  \leq
C\left\|  x-y\right\|  ,\text{ \ }x,y\in\mathfrak{P}\left(  \Omega\right)  ,
\]
and $p\left(  0\right)  =q$ is in the interior of $E$, then there exists a
unique global solution to the ordinary differential equation: $\frac{\text{d}%
}{\text{d}t}p\left(  t\right)  =p\left(  t\right)  \Gamma\left(  p\left(
t\right)  \right)  $ with $p\left(  0\right)  =q$, within $E$.

To deduce whether the $\Gamma\left(  x\right)  $ is (locally) Lipschitz on a
set $E\subseteq\mathfrak{P}\left(  \Omega\right)  $, Li et al. \cite{LiD:2014}
and Li and Lui \cite{LiL:2014} gave an algorithmic method through dealing with
some matrices of infinite orders.

\textbf{(b) The limiting processes}

To discuss the limit: For $t\geq0$, $\mu^{N}\left(  t\right)  \Longrightarrow
p\left(  t\right)  $ as $N\rightarrow\infty$, we need to set up some suitable
conditions in order to guarantee the existence of such a limit.

Let $e_{k,j}$ be the unit vector of infinite dimension in which the $\left(
k,j\right)  $th entry is one and all the others are zero. Note that the
empirical measure process $\mu^{\left(  N\right)  }=\left\{  \mu^{N}\left(
t\right)  :t\geq0\right\}  $ is a Markov process on the state space
$\mathfrak{P}_{N}\left(  \Omega\right)  $ where $\mathfrak{P}_{N}\left(
\Omega\right)  =\mathfrak{P}\left(  \Omega\right)  \cap\left(  \frac{1}%
{N}\Omega\right)  $, the possible jumps of $\mu^{\left(  N\right)  }$ are of
the form $\left(  e_{l,i}-e_{k,j}\right)  /N$ for $\left(  l,i\right)
\neq\left(  k,j\right)  $, and $\left(  k,j\right)  ,\left(  l,i\right)
\in\Omega$. If $\mu^{N}\left(  t\right)  =x\in\mathfrak{P}_{N}\left(
\Omega\right)  $ at time $t\geq0$, then $Nx_{k,j}$ denotes State $\left(
k,j\right)  $ of the big network. Hence the transition rate of the Markov
process from State $\left(  k,j\right)  $ to State $\left(  l,i\right)  $,
corresponding to the tagged big network, is given by $Nx_{k,j}\Gamma
_{k,j;l,i}^{\left(  N\right)  }\left(  x\right)  $. Based on this, the
generator $\mathbf{A}^{\left(  N\right)  }$ of the Markov process
$\mu^{\left(  N\right)  }$ is given by%
\[
\mathbf{A}^{\left(  N\right)  }f\left(  x\right)  =\sum_{\left(  k,j\right)
\in\Omega}\sum_{\substack{\left(  l,i\right)  \in\Omega\\\left(  l,i\right)
\neq\left(  k,j\right)  }}Nx_{k,j}\Gamma_{k,j;l,i}^{\left(  N\right)  }\left(
x\right)  \left[  f\left(  x+\frac{1}{N}\left(  e_{l,i}-e_{k,j}\right)
\right)  -f\left(  x\right)  \right]  ,
\]
where $f\left(  x\right)  $ is a real function on
$\mathfrak{P}_{N}\left( \Omega\right)  $, and there are two types of
boundary states: That a task enters the big network corresponds to
an arriving boundary state: $\left( k,j\right)  =\left(  0,j\right)
$; while a task is completed and immediately leaves the big network
corresponds to a departed boundary state: $\left( l,i\right)
=\left(  0,i\right)  $. It is easy to see that as $N\rightarrow
\infty$
\[
\mathbf{A}^{\left(  N\right)  }f\left(  x\right)  \rightarrow\sum_{\left(
k,j\right)  \in\Omega}\sum_{\substack{\left(  l,i\right)  \in\Omega\\\left(
l,i\right)  \neq\left(  k,j\right)  }}x_{k,j}\Gamma_{k,j;l,i}\left(  x\right)
\left[  \frac{\partial}{\partial x_{l,i}}f\left(  x\right)  -\frac{\partial
}{\partial x_{k,j}}f\left(  x\right)  \right]  \overset{\text{def}}%
{=}\mathbf{A}f\left(  x\right)  .
\]

\begin{The}
Suppose that for $\left(  k,j\right)  ,\left(  l,i\right)  \in\Omega$ with
$\left(  k,j\right)  \neq\left(  l,i\right)  $, there exists a Lipschitz
continuous function $\Gamma_{k,j;l,i}\left(  p\right)  :\mathfrak{P}\left(
\Omega\right)  \rightarrow\lbrack0,+\infty)$ such that $\Gamma_{k,j;l,i}%
^{\left(  N\right)  }\left(  p\right)  \rightarrow\Gamma_{k,j;l,i}\left(
p\right)  $ uniformly on $\mathfrak{P}\left(  \Omega\right)  $. If $\left\{
\mu^{\left(  N\right)  }\left(  0\right)  \right\}  $ converges in probability
to $q\in\mathfrak{P}\left(  \Omega\right)  $, then $\left\{  \mu^{\left(
N\right)  }\left(  t\right)  \right\}  $ converges uniformly on compact time
intervals in probability to $p\left(  t\right)  \in\mathfrak{P}\left(
\Omega\right)  $ for $t\geq0$, where the probability vector $p\left(
t\right)  $ is the unique global solution to the ordinary differential
equation: $\frac{\text{d}}{\text{d}t}p\left(  t\right)  =p\left(  t\right)
\Gamma\left(  p\left(  t\right)  \right)  $ with $p\left(  0\right)  =q$.
\end{The}

\textbf{Proof:} The proof may directly follow from Theorem 2.11 in Kurtz
\cite{Kur:1970}. Here, we only give a simple outline as follows. Firstly,
notice that%
\[
F^{\left(  N\right)  }\left(  p\right)  =\sum_{\left(  k,j\right)  ,\left(
l,i\right)  \in\Omega}Np_{k,j}\left(  \frac{1}{N}e_{l,i}-\frac{1}{N}%
e_{k,j}\right)  \Gamma_{k,j;l,i}^{\left(  N\right)  }\left(  p\right)
\]
and%
\[
F\left(  p\right)  =\sum_{\left(  k,j\right)  ,\left(  l,i\right)  \in\Omega
}p_{k,j}\left(  e_{l,i}-e_{k,j}\right)  \Gamma_{k,j;l,i}\left(  p\right)  ,
\]
where as $N\rightarrow\infty$%
\[
\Gamma_{k,j;l,i}^{\left(  N\right)  }\left(  p\right)  \rightarrow
\Gamma_{k,j;l,i}\left(  p\right)  ,
\]
and $\Gamma\left(  x\right)  $ is (locally) Lipschitz on a set $E\subseteq
\mathfrak{P}\left(  \Omega\right)  $, thus for the sequence $\left\{
\mu^{(N)}(t),t\geq0\right\}  $ of Markov processes, it follows from Equation
(III.10.13) in Rogers and Williams \cite{Rog:1994} or Page 162 in Ethier and
Kurtz \cite{Eth:1986}\ that%
\[
\mathbf{M}^{\left(  N\right)  }\left(  t\right)  =\mu^{(N)}(t)-\mu
^{(N)}(0)-\int_{0}^{t}\mu^{(N)}(x)\Gamma\left(  \mu^{(N)}(x)\right)
\text{d}x
\]
is a martingale with respect to each $N\geq1$. Therefore, if $\left\{
\mu^{(N)}(0)\right\}  $\ converges weakly to$\ q\in\mathfrak{P}\left(
\Omega\right)  $ as $N\rightarrow\infty$, then $\left\{  \mu^{(N)}%
(t),N\geq1\right\}  $ converges weakly in $D_{\mathcal{F}}[0,+\infty)$ endowed
with the Skorohod topology to the solution $p\left(  t\right)  $ to the
ordinary differential equation: $\frac{\text{d}}{\text{d}t}p\left(  t\right)
=p\left(  t\right)  \Gamma\left(  p\left(  t\right)  \right)  $ with $p\left(
0\right)  =q$, within $\mathfrak{P}\left(  \Omega\right)  $. This completes
the proof. \textbf{{\rule{0.08in}{0.08in}}}

\begin{Rem}
Benaim and Le Boudec \cite{Ben:2008} applied the mean-field theory to studying
the discrete-time system of $N$ weakly interacting objects, and for
$t=0,1,2,\ldots$, the states of this entire system are expressed as
$\mathbf{Y}^{\left(  N\right)  }\left(  t\right)  =\left(  X_{1}^{\left(
N\right)  }\left(  t\right)  ,X_{2}^{\left(  N\right)  }\left(  t\right)
,\ldots,X_{N}^{\left(  N\right)  }\left(  t\right)  ;R^{\left(  N\right)
}\left(  t\right)  \right)  $ where $X_{k}^{\left(  N\right)  }\left(
t\right)  $ is the state of the $k$th object for $1\leq k\leq N$, and
$R^{\left(  N\right)  }\left(  t\right)  $ is the state of the common resource
of the $N$ objects. Differently from that of Benaim and Le Boudec
\cite{Ben:2008}, this paper uses the mean-field theory to studying the
continuous-time system of $N$ weakly interacting objects, where the states of
each object is described as a block-structured Markov process (i.e., a Markov
process under a stochastic enviornment $J^{\left(  N\right)  }\left(
t\right)  $), and it is easy to see that the stochastic enviornment
$J^{\left(  N\right)  }\left(  t\right)  $ may be regarded as the resource
$R^{\left(  N\right)  }\left(  t\right)  $. On the other hand, we provide a
simple method to describe the weak interaction of the $N$ objects through
introduction to the matrix $\Gamma^{\left(  N\right)  }\left(  \mu^{N}\left(
t\right)  \right)  $ where $\mu^{N}\left(  t\right)  $ is the empirical
measure of the system of the $N$ weakly interacting objects. Although our
method is simple to deal with the weak convergence $\Gamma^{\left(  N\right)
}\left(  \mu^{N}\left(  t\right)  \right)  \Longrightarrow\Gamma\left(
p\left(  t\right)  \right)  $ as $N\rightarrow\infty$, it is useful and
effective in the mean-field study of many continuous-time big networks,
readers may refer to Li et al. \cite{LiD:2014, LiD:2015} and Li
\cite{LiL:2014} for more details.
\end{Rem}

\section{The Fixed Points}

In this section, we use the UL-type $RG$-factorization to provide some
effective algorithms for computing the fixed points of the ordinary
differential equation: $\frac{\text{d}}{\text{d}t}p\left(  t\right)  =p\left(
t\right)  \Gamma\left(  p\left(  t\right)  \right)  $ with $p\left(  0\right)
=q$. Further, we set up a nonlinear characteristic equation of the censoring
matrix to level $0$, which is satisfied by the fixed points.

A point $\pi\in$ $\mathfrak{P}\left(  \Omega\right)  $ is said to be a fixed
point of the ordinary differential equation: $\frac{\text{d}}{\text{d}%
t}p\left(  t\right)  =p\left(  t\right)  \Gamma\left(  p\left(  t\right)
\right)  $ with $p\left(  0\right)  =q$, if $p\left(  t\right)  \rightarrow
\pi$ as $t\rightarrow+\infty$, and
\[
\lim_{t\rightarrow+\infty}\left[  \frac{\text{d}}{\text{d}t}p\left(  t\right)
\right]  =0.
\]
In this case, it is clear that%
\begin{equation}
\pi\Gamma\left(  \pi\right)  =0, \label{FixedPE}%
\end{equation}
which is an infinite-dimensional system of nonlinear equations. In general,
there are more difficulties and challenging due to both the infinite order of
and the nonlinear structure of the matrix $\Gamma\left(  \pi\right)  $ when
solving the fixed point equation (\ref{FixedPE}) together with $\pi e=1$,
where $e$ is a column vector of ones with a suitable size.

It is easy to check that for every $\pi\in$ $\mathfrak{P}\left(
\Omega\right)  $, $\Gamma\left(  \pi\right)  $ is the infinitesimal generator
of an irreducible continuous-time Markov process. Based on Li \cite{Li:2010},
we can develop the UL-type $RG$-factorization of the matrix $\Gamma\left(
\pi\right)  $. To that end, we partition the matrix $\Gamma\left(  \pi\right)
$ as
\[
\Gamma\left(  \pi\right)  =\left(
\begin{array}
[c]{cc}%
T\left(  \pi\right)  & U\left(  \pi\right) \\
V\left(  \pi\right)  & W\left(  \pi\right)
\end{array}
\right)
\]
according to the level sets $L_{\leq n}$ and $L_{\geq n+1}$ for $n\geq0$.
Since the Markov chain $\Gamma\left(  \pi\right)  $ is irreducible, it is
clear that the two truncated chains with infinitesimal generators $T\left(
\pi\right)  $ and $W\left(  \pi\right)  $ are all transient, and also the
matrices $T\left(  \pi\right)  $ and $W\left(  \pi\right)  $ are all
invertible from a different understanding that the inverse of the matrix
$T\left(  \pi\right)  $ is ordinary, but the invertibility of the matrix
$W\left(  \pi\right)  $ is different under an infinite-dimensional meaning.
Although the matrix $W\left(  \pi\right)  $ of infinite size may have multiple
inverses, we in general are interested in the maximal non-positive inverse
$W_{\max}^{-1}\left(  \pi\right)  $ of $W\left(  \pi\right)  $, i.e.,
$W^{-1}\left(  \pi\right)  \leq W_{\max}^{-1}\left(  \pi\right)  \leq0$ for
any non-positive inverse $W^{-1}\left(  \pi\right)  $. Of course,
$0\leq\left[  -W\left(  \pi\right)  \right]  _{\min}^{-1}\leq\left[  -W\left(
\pi\right)  \right]  ^{-1}$ for any nonnegative inverse $\left[  -W\left(
\pi\right)  \right]  ^{-1}$ of $-W\left(  \pi\right)  $, that is, $\left[
-W\left(  \pi\right)  \right]  _{\min}^{-1}$ is the minimal nonnegative
inverse of $-W\left(  \pi\right)  $. Based on this, for $n\geq0$ we write%
\[
\Gamma^{\left[  \leq n\right]  }\left(  \pi\right)  =T\left(  \pi\right)
+U\left(  \pi\right)  \left[  -W\left(  \pi\right)  \right]  _{\min}%
^{-1}V\left(  \pi\right)  =\left(
\begin{array}
[c]{cccc}%
\phi_{0,0}^{\left(  n\right)  }\left(  \pi\right)  & \phi_{0,1}^{\left(
n\right)  }\left(  \pi\right)  & \cdots & \phi_{0,n}^{\left(  n\right)
}\left(  \pi\right) \\
\phi_{1,0}^{\left(  n\right)  }\left(  \pi\right)  & \phi_{1,1}^{\left(
n\right)  }\left(  \pi\right)  & \cdots & \phi_{1,n}^{\left(  n\right)
}\left(  \pi\right) \\
\vdots & \vdots &  & \vdots\\
\phi_{n,0}^{\left(  n\right)  }\left(  \pi\right)  & \phi_{n,1}^{\left(
n\right)  }\left(  \pi\right)  & \cdots & \phi_{n,n}^{\left(  n\right)
}\left(  \pi\right)
\end{array}
\right)  ,
\]
where the size of the matrix $\phi_{j,j}^{\left(  n\right)  }\left(
\pi\right)  $ is $m_{j}$ for $0\leq j\leq n$, and the sizes of other matrices
can be determined accordingly. It is clear from Section 7 of Chapter 2 in Li
\cite{Li:2010} that for $n\geq0$, $0\leq i$, $j\leq n$,%
\[
\phi_{i,j}^{\left(  n\right)  }\left(  \pi\right)  =\Gamma_{i,j}\left(
\pi\right)  +\sum\limits_{k=n+1}^{\infty}\phi_{i,k}^{\left(  k\right)
}\left(  \pi\right)  \left[  -\phi_{k,k}^{\left(  k\right)  }\left(
\pi\right)  \right]  ^{-1}\phi_{k,j}^{\left(  k\right)  }\left(  \pi\right)
.
\]
Let%
\[
\Psi_{n}\left(  \pi\right)  =\phi_{n,n}^{\left(  n\right)  }\left(
\pi\right)  ,\text{ \ }n\geq0;
\]%
\[
R_{i,j}\left(  \pi\right)  =\phi_{i,j}^{\left(  j\right)  }\left(  \pi\right)
\left[  -\phi_{j,j}^{\left(  k\right)  }\left(  \pi\right)  \right]
^{-1},\text{ \ }0\leq i<j;
\]
and%
\[
G_{i,j}\left(  \pi\right)  =\left[  -\phi_{i,i}^{\left(  k\right)  }\left(
\pi\right)  \right]  ^{-1}\phi_{i,j}^{\left(  i\right)  }\left(  \pi\right)
,\text{ \ }0\leq j<i.
\]
Then the UL-type $RG$-factorization of the matrix $\Gamma\left(  \pi\right)  $
is given by
\begin{equation}
\Gamma\left(  \pi\right)  =\left[  I-R_{U}\left(  \pi\right)  \right]
\Psi_{D}\left(  \pi\right)  \left[  I-G_{L}\left(  \pi\right)  \right]  ,
\label{factor}%
\end{equation}
where%
\[
R_{U}\left(  \pi\right)  =\left(
\begin{array}
[c]{ccccc}%
0 & R_{0,1}\left(  \pi\right)  & R_{0,2}\left(  \pi\right)  & R_{0,3}\left(
\pi\right)  & \cdots\\
& 0 & R_{1,2}\left(  \pi\right)  & R_{1,3}\left(  \pi\right)  & \cdots\\
&  & 0 & R_{2,3}\left(  \pi\right)  & \cdots\\
&  &  & 0 & \cdots\\
&  &  &  & \ddots
\end{array}
\right)  ,
\]%
\[
\Psi_{D}\left(  \pi\right)  =\hbox{diag}  \left(  \Psi_{0}\left(  \pi\right)
,\Psi_{1}\left(  \pi\right)  ,\Psi_{2}\left(  \pi\right)  ,\Psi_{3}\left(
\pi\right)  ,\ldots\right)
\]
and%
\[
G_{L}\left(  \pi\right)  =\left(
\begin{array}
[c]{ccccc}%
0 &  &  &  & \\
G_{1,0}\left(  \pi\right)  & 0 &  &  & \\
G_{2,0}\left(  \pi\right)  & G_{2,1}\left(  \pi\right)  & 0 &  & \\
G_{3,0}\left(  \pi\right)  & G_{3,1}\left(  \pi\right)  & G_{3,2}\left(
\pi\right)  & 0 & \\
\vdots & \vdots & \vdots & \vdots & \ddots
\end{array}
\right)  .
\]

Based on the UL-type $RG$-factorization (\ref{factor}), it follows from
Subsection 2.7.3 in Li \cite{Li:2010} that the fixed point $\pi$ is given by%
\begin{equation}
\left\{
\begin{array}
[c]{l}%
\pi_{0}=\tau x_{0}\left(  \pi\right)  ,\\
\pi_{k}=\sum\limits_{i=0}^{k-1}\pi_{i}R_{i,k}\left(  \pi\right)  ,\text{
\ }k\geq1,
\end{array}
\right.  \label{FixedP}%
\end{equation}
where $x_{0}\left(  \pi\right)  $ is the stationary probability vector of the
censored Markov chain $\Psi_{0}\left(  \pi\right)  $ to level $0$,\ and the
scalar $\tau$ is determined by $\sum_{k=0}^{\infty}\pi_{k}e=1$ uniquely.

Using the expression (\ref{FixedP}) of the fixed point $\pi$, we set up an
important relation as follows:%
\begin{equation}
\pi=\left(  \tau x_{0}\left(  \pi\right)  ,\pi_{0}R_{0,1}\left(  \pi\right)
,\sum\limits_{i=0}^{1}\pi_{i}R_{i,k}\left(  \pi\right)  ,\sum\limits_{i=0}%
^{2}\pi_{i}R_{i,k}\left(  \pi\right)  ,\ldots\right)  , \label{FixedPE-1}%
\end{equation}
which is called a fixed point equation with $R$-measure.

In what follows we consider two special cases in order to further explain the
fixed point equation (\ref{FixedPE-1}) with $R$-measure.

\textbf{Case one: Nonlinear Markov processes of GI/M/1 type}

In this case, the infinitesimal generator $\Gamma\left(  \pi\right)  $ is
given by%
\[
\Gamma\left(  \pi\right)  =\left(
\begin{array}
[c]{ccccc}%
B_{1}\left(  \pi\right)  & B_{0}\left(  \pi\right)  &  &  & \\
B_{2}\left(  \pi\right)  & A_{1}\left(  \pi\right)  & A_{0}\left(  \pi\right)
&  & \\
B_{3}\left(  \pi\right)  & A_{2}\left(  \pi\right)  & A_{1}\left(  \pi\right)
& A_{0}\left(  \pi\right)  & \\
\vdots & \vdots & \vdots & \vdots & \ddots
\end{array}
\right)  .
\]
Let $R\left(  \pi\right)  $ be the minimal nonnegative solution to the
nonlinear matrix equation%
\[
\sum_{k=0}^{\infty}R^{k}\left(  \pi\right)  A_{k}\left(  \pi\right)  =0.
\]
Then%
\[
\pi_{k}=\pi_{1}R^{k-1}\left(  \pi\right)  ,\text{ \ }k\geq1,
\]
where the two vectors $\pi_{0}$ and $\pi_{1}$ satisfy the following system of
nonlinear matrix equations%
\[
\left(  \pi_{0},\pi_{1}\right)  \left(
\begin{array}
[c]{cc}%
B_{1}\left(  \pi\right)  & B_{0}\left(  \pi\right) \\
\sum\limits_{k=0}^{\infty}R^{k}\left(  \pi\right)  B_{k+2}\left(  \pi\right)
& \sum\limits_{k=0}^{\infty}R^{k}\left(  \pi\right)  A_{k+1}\left(
\pi\right)
\end{array}
\right)  =0
\]
and%
\[
\pi_{0}e+\pi_{1}\left[  I-R\left(  \pi\right)  \right]  ^{-1}e=1.
\]
Thus, the fixed point equation (\ref{FixedPE-1}) with $R$-measure is
simplified as%
\[
\pi=\left(  \pi_{0},\pi_{1},\pi_{1}R\left(  \pi\right)  ,\pi_{1}R^{2}\left(
\pi\right)  ,\ldots\right)  .
\]

\textbf{Case two: Nonlinear Markov processes of M/G/1 type}

In this case, the infinitesimal generator $\Gamma\left(  \pi\right)  $ is
given by%
\[
\Gamma\left(  \pi\right)  =\left(
\begin{array}
[c]{ccccc}%
B_{1}\left(  \pi\right)  & B_{2}\left(  \pi\right)  & B_{3}\left(  \pi\right)
& B_{4}\left(  \pi\right)  & \cdots\\
B_{0}\left(  \pi\right)  & A_{1}\left(  \pi\right)  & A_{2}\left(  \pi\right)
& A_{3}\left(  \pi\right)  & \cdots\\
& A_{0}\left(  \pi\right)  & A_{1}\left(  \pi\right)  & A_{2}\left(
\pi\right)  & \cdots\\
&  & A_{0}\left(  \pi\right)  & A_{1}\left(  \pi\right)  & \cdots\\
&  &  & \ddots & \ddots
\end{array}
\right)  .
\]
Let $G\left(  \pi\right)  $ be the minimal nonnegative solution to the
nonlinear matrix equation%
\[
\sum_{k=0}^{\infty}A_{k}\left(  \pi\right)  G^{k}\left(  \pi\right)  =0.
\]
Then%
\[
\Psi_{0}\left(  \pi\right)  =B_{1}\left(  \pi\right)  +\sum_{k=2}^{\infty
}B_{k}\left(  \pi\right)  G^{k-2}\left(  \pi\right)  G_{1}\left(  \pi\right)
,
\]%
\[
\Psi\left(  \pi\right)  =A_{1}\left(  \pi\right)  +\sum_{k=2}^{\infty}%
A_{k}\left(  \pi\right)  G^{k-1}\left(  \pi\right)  ;
\]
and the $R$-measure%
\[
R_{0,j}\left(  \pi\right)  =\left[  \sum_{k=j+1}^{\infty}B_{k}\left(
\pi\right)  G^{k-1}\left(  \pi\right)  \right]  \left[  -\Psi\left(
\pi\right)  \right]  ^{-1},\text{ \ }j\geq1,
\]
for $i\geq1$%
\[
R_{j}\left(  \pi\right)  =\left[  \sum_{k=j+1}^{\infty}A_{k}\left(
\pi\right)  G^{k-1}\left(  \pi\right)  \right]  \left[  -\Psi\left(
\pi\right)  \right]  ^{-1},\text{ \ }j\geq1.
\]
The fixed point $\pi$ is given by%
\[
\left\{
\begin{array}
[c]{l}%
\pi_{0}=\tau x_{0}\left(  \pi\right)  ,\\
\pi_{k}=\pi_{0}R_{0,k}\left(  \pi\right)  +\sum\limits_{i=1}^{k-1}\pi
_{i}R_{k-i}\left(  \pi\right)  ,\ \ k\geq1,
\end{array}
\right.
\]
where $x_{0}\left(  \pi\right)  $ is the stationary probability vector of the
censored Markov chain $\Psi_{0}\left(  \pi\right)  $ to level $0$,\ and the
scalar $\tau$ is determined by $\sum_{k=0}^{\infty}\pi_{k}e=1$ uniquely. Thus,
the fixed point equation (\ref{FixedPE-1}) with $R$-measure is simplified as%
\[
\pi=\left(  \tau x_{0}\left(  \pi\right)  ,\pi_{0}R_{0,1}\left(  \pi\right)
,\pi_{0}R_{0,2}\left(  \pi\right)  +\pi_{1}R_{1}\left(  \pi\right)  ,\pi
_{0}R_{0,3}\left(  \pi\right)  +\sum\limits_{i=1}^{2}\pi_{i}R_{k-i}\left(
\pi\right)  ,\ldots\right)  .
\]

Now, we write the fixed point equation (\ref{FixedPE-1}) with $R$-measure as a
functional form: $\pi=\mathbf{F}\left(  \mathfrak{R}\left(  \pi\right)
\right)  $ where $\mathfrak{R}\left(  \pi\right)  $ is related to the
$R$-measure, as shown in the above two special cases. Based on this, we can
provide an approximative algorithm as follows:

\textbf{Algorithm I: Computation of the fixed points}

\textbf{Step one:} Taking any initial probability vector: $\pi^{\left(
0\right)  }\in\mathfrak{P}\left(  \Omega\right)  $.

\textbf{Step two:} Computing the infinitesimal generator: $\Gamma\left(
\pi^{\left(  0\right)  }\right)  $; and then compute the $R$-measure, which
gives $\pi^{\left(  1\right)  }=\mathbf{F}\left(  \mathfrak{R}\left(
\pi^{\left(  0\right)  }\right)  \right)  $.

\textbf{Step three:} For $N\geq2$, compute $\pi^{\left(  N+1\right)
}=\mathbf{F}\left(  \mathfrak{R}\left(  \pi^{\left(  N\right)  }\right)
\right)  $.

\textbf{Step four:} For a sufficiently small $\varepsilon>0$, if $\left\|
\pi^{\left(  N+1\right)  }-\pi^{\left(  N\right)  }\right\|  <\varepsilon$,
then the computation is over; otherwise we go to Step three.

Note that it is possible for some practical big networks that there exist
multiple fixed points because the infinitesimal generator $\Gamma\left(
\pi\right)  $ is more general, e.g., see such examples given in Marbukh
\cite{Mar:1984}, Gibbens at al. \cite{Gib:1990}, Kelly \cite{Kel:1991} (see
Page 349), Dobrushin \cite{Dob:1993}, Antunes et al. \cite{Ant:2008} and Tibi
\cite{Tib:2010}. In this case, it is a key to design a suitable initial
probability vector: $\pi^{\left(  0\right)  }\in\mathfrak{P}\left(
\Omega\right)  $, for example:

(1) For any given integer $m\geq1$, we take%
\[
\pi^{\left(  0\right)  }=\left(  \frac{1}{m},\frac{1}{m},\ldots,\frac{1}%
{m},0,\ldots\right)  .
\]

(2) For any given parameter $\rho\in\left(  0,1\right)  $, we take%
\[
\pi^{\left(  0\right)  }=\left(  1-\rho,\rho\left(  1-\rho\right)  ,\rho
^{2}\left(  1-\rho\right)  ,\rho^{3}\left(  1-\rho\right)  ,\ldots\right)  .
\]

(3) For any given parameter $\lambda>0$, we take%
\[
\pi^{\left(  0\right)  }=\left(  e^{-\lambda},e^{-\lambda}\frac{\lambda}%
{1!},e^{-\lambda}\frac{\lambda^{2}}{2!},e^{-\lambda}\frac{\lambda^{3}}%
{3!},\ldots\right)  .
\]

(4) It is obvious that we can construct a suitable vector $\pi^{\left(
0\right)  }$ through some practical observation and experience, for instance,
a discrete-time PH distribution of order 2 is established by means of only the
first three moments of a random variable.

Now, we provide another algorithm for computing the fixed points. To do end,
we set up a characteristic equation of the censoring matrix $\Psi_{0}\left(
\pi\right)  $ to level $0$, while the characteristic equation is satisfied by
the fixed points.

Note that for the censored Markov chain $\Psi_{0}\left(  \pi\right)  $ to
level $0$, we have%
\[
x_{0}\left(  \pi\right)  \Psi_{0}\left(  \pi\right)  =0,\text{ \ }x_{0}\left(
\pi\right)  e=1;
\]%
\[
\pi_{0}=\tau x_{0}\left(  \pi\right)  ,\text{ \ }\pi_{0}e=\tau\in\left(
0,1\right)  .
\]
Thus it is easy to see from the irreducibility of the matrix $\Gamma\left(
\pi\right)  $ that the matrix $\Psi_{0}\left(  \pi\right)  $ of size $m_{0}$
is also irreducible, hence this gives rank$\left(  \Psi_{0}\left(  \pi\right)
\right)  =m_{0}-1$. For the $\mathbf{M}$ matrix $\Psi_{0}\left(  \pi\right)
$, its eigenvalue with the maximal real part is equal to zero due to $\Psi
_{0}\left(  \pi\right)  e=0$. Let the characteristic equation be $f_{x}\left(
\pi\right)  =\det\left(  xI-\Psi_{0}\left(  \pi\right)  \right)  =0$. Then the
fixed points satisfy the characteristic equation $f_{0}\left(  \pi\right)
=\det\left(  -\Psi_{0}\left(  \pi\right)  \right)  =0$, and hence $\det\left(
\Psi_{0}\left(  \pi\right)  \right)  =0$. Hence the fixed points satisfy the
system of nonlinear equations as follows:%
\begin{equation}
\left\{
\begin{array}
[c]{l}%
\det\left(  \Psi_{0}\left(  \pi\right)  \right)  =0,\\
\text{rank}\left(  \Psi_{0}\left(  \pi\right)  \right)  =m_{0}-1.
\end{array}
\right.  \label{CharF}%
\end{equation}

Note that (\ref{CharF}) provide another algorithm for computing the fixed
points as follows:

\textbf{Algorithm II: Computation of the fixed points}

\textbf{Step one:} Providing a numerical solution $\widehat{\pi}$ to the
nonlinear characteristic equation: $\det\left(  \Psi_{0}\left(  \pi\right)
\right)  =0$.

\textbf{Step two:} Check whether rank$\left(  \Psi_{0}\left(  \widehat{\pi
}\right)  \right)  =m_{0}-1$. If Yes, then $\widehat{\pi}$ is a fixed point.
If No, then going to Step one.

\section{Stability and Metastability}

In this section, we first discuss the Birkhoff center and the locally stable
fixed points of the dynamic system of mean-field equations: $\frac{\text{d}%
}{\text{d}t}p\left(  t\right)  =p\left(  t\right)  \Gamma\left(  p\left(
t\right)  \right)  $ with $p\left(  0\right)  =q$. Then we apply the Lyapunov
functions and the relative entropy to studying the stability or metastability
of the big networks. Furthermore, we provide several interesting open problems
with detailed interpretation.

We write%
\[
\mathbb{S}=\left\{  \pi:\pi\Gamma\left(  \pi\right)  =0,\text{ }\pi
e=1\right\}  .
\]
Then it is clear that%
\[
\mathbb{S}=\left\{  \pi:\pi=\left(  \tau x_{0}\left(  \pi\right)  ,\pi
_{0}R_{0,1}\left(  \pi\right)  ,\sum\limits_{i=0}^{1}\pi_{i}R_{i,k}\left(
\pi\right)  ,\sum\limits_{i=0}^{2}\pi_{i}R_{i,k}\left(  \pi\right)
,\ldots\right)  ,\text{ }\pi e=1\right\}
\]
or%
\[
\mathbb{S}=\left\{  \pi:\det\left(  \Psi_{0}\left(  \pi\right)  \right)
=0,\text{rank}\left(  \Psi_{0}\left(  \pi\right)  \right)  =m_{0}-1,\text{
}\pi e=1\right\}
\]
with $\pi_{0}=\tau x_{0}\left(  \pi\right)  $ and $\pi_{k}=\sum\limits_{i=0}%
^{k-1}\pi_{i}R_{i,k}\left(  \pi\right)  $ for $k\geq1$.

Since the vector equation $\pi\Gamma\left(  \pi\right)  =0$, together with
$\pi e=1$, is nonlinear, it is possible for some practical big networks that
there exist multiple elements in the set $\mathbb{S}$. Such practical examples
include Marbukh \cite{Mar:1984}, Gibbens at al. \cite{Gib:1990}, Kelly
\cite{Kel:1991} (see Page 349), Dobrushin \cite{Dob:1993}, Antunes et al.
\cite{Ant:2008} and Tibi \cite{Tib:2010}. At the same time, an argument by
analytic functions can indicate an important result: The elements of the set
$\mathbb{S}$ are isolated.

To describe the isolated element structure of the set $\mathbb{S}$, we often
need to use the Birkhoff center of the dynamic system of mean-field equations.
Notice that the Birkhoff center is used to check whether the fixed point is
unique or not. Based on this, our discussion includes the following two cases:

\textbf{Case one: }$N\rightarrow\infty$. In this case, we denote by
$\Phi\left(  t\right)  $ a solution to the system of differential equations
$\frac{\text{d}}{\text{d}t}p\left(  t\right)  =p\left(  t\right)
\Gamma\left(  p\left(  t\right)  \right)  $ with $p\left(  0\right)  =q$.
Thus, the Birkhoff center of the solution $\Phi\left(  t\right)  $ is defined
as%
\begin{align*}
\mathbf{\Theta}=  &  \left\{  \overline{P}\in\mathfrak{P}\left(
\Omega\right)  :\overline{P}=\lim_{k\rightarrow\infty}\Phi\left(
t_{k}\right)  \text{ for any scale sequence}\right. \\
&  \left.  \left\{  t_{k}\right\}  \text{ with }t_{l}\geq0\text{ for }%
l\geq1\text{ and }\lim_{k\rightarrow\infty}t_{k}=+\infty\right\}  .
\end{align*}
Notice that perhaps $\mathbf{\Theta}$ contains the limit cycles or the
equilibrium points (the local minimal points, or the local maximal points, or
the saddle points). Thus it is clear that $\mathbb{S}\subset\mathbf{\Theta}$.
Obviously, the limiting empirical Markov process $\left\{  \mathbf{Y}\left(
t\right)  :t\geq0\right\}  $ spends most of its time in the Birkhoff center
$\mathbf{\Theta}$, where $\mathbf{Y}\left(  t\right)  =\lim_{N\rightarrow
\infty}\mu^{N}\left(  t\right)  $ weakly.

\textbf{Case two: }$t\rightarrow+\infty$. In this case, we write%
\[
\pi^{\left(  N\right)  }=\lim_{t\rightarrow+\infty}\mu^{N}\left(  t\right)
,\text{ \ weakly},
\]
if for each $N=1,2,3,\ldots$, the system of $N$ weakly interacting big
networks is stable.

Let%
\begin{align*}
\Xi=  &  \left\{  \overline{\pi}\in\mathfrak{P}\left(  \Omega\right)
:\overline{\pi}=\lim_{k\rightarrow\infty}\pi^{\left(  N_{k}\right)  }\text{
for any positive integer sequence}\right. \\
&  \left.  \left\{  N_{k}\right\}  \text{ with }1\leq N_{1}\leq N_{2}\leq
N_{3}\leq\cdots\text{ and }\lim_{k\rightarrow\infty}N_{k}=\infty\right\}  .
\end{align*}
It is easy to see that%
\[
\mathbb{S}\subset\Xi\subset\mathbf{\Theta.}%
\]
At the same time, it is clear that%
\begin{align*}
\mathbb{S}=  &  \left\{  \text{the local minimal points in }\mathbf{\Theta
}\right\}  \cup\left\{  \text{the local maximal points in }\mathbf{\Theta
}\right\} \\
&  \cup\left\{  \text{the saddle points in }\mathbf{\Theta}\right\}  .
\end{align*}
and%
\[
\mathbf{\Theta-}\mathbb{S}=\left\{  \text{the limit cycles in }\mathbf{\Theta
}\right\}  .
\]

In what follows, we discuss stability or metastability of the big networks.

To analyze the stability or metastability, a key is to determine a Lyapunov
function for the dynamic system of mean-field equations: $\frac{\text{d}%
}{\text{d}t}p\left(  t\right)  =p\left(  t\right)  \Gamma\left(  p\left(
t\right)  \right)  $ with $p\left(  0\right)  =q$. The Lyapunov function $g$
defined on $\mathfrak{P}\left(  \Omega\right)  $ is constructed such that%
\begin{equation}
y\Gamma\left(  y\right)  \cdot\nabla g\left(  y\right)  \leq0,\text{ \ }%
y\in\mathfrak{P}\left(  \Omega\right)  . \label{LyaF-1}%
\end{equation}
It is easy to see that if $\pi\in\mathbb{S}$, then $\pi\Gamma\left(
\pi\right)  \cdot\nabla g\left(  \pi\right)  =0$ due to the fact that
$\pi\Gamma\left(  \pi\right)  =0$. On the other hand, if $\pi\Gamma\left(
\pi\right)  \cdot\nabla g\left(  \pi\right)  =0$, then $\pi\in\mathbb{S}$.

Let $\left|  \mathbb{S}\right|  $ be the number of elements in the set
$\mathbb{S}$. If $\left|  \mathbb{S}\right|  =1$, then%
\[
\lim_{N\rightarrow\infty}\lim_{t\rightarrow+\infty}\mu^{N}\left(  t\right)
=\lim_{t\rightarrow+\infty}\lim_{N\rightarrow\infty}\mu^{N}\left(  t\right)
=\pi,\text{ \ weakly}.
\]

If $\left|  \mathbb{S}\right|  \geq2$, then the system of big networks
exhibits a metastability property, that is, the evolutionary process of the
tagged big network with a mean-field modification switches from one stable
point to the other one after a long residence time. In the study of
metastability, it is a key to estimate the expected value of such a residence
time. See Bovier \cite{Bov:2006}, and Olivieri and Vares \cite{Oli:2005} for
more details.

An interesting issue in the study of big networks is to analyze stability or
metastability of the corresponding nonlinear Markov processes. On this line,
it is a key to construct a Lyapunov function or a local Lyapunov function.
Note that the relative entropy function in some sense can define a globally
attracting Lyapunov function.

For $p,q\in\mathfrak{P}\left(  \Omega\right)  $, we define the relative
entropy of $p$ with respect to $q$ as%
\[
R\left(  p||q\right)  =\sum_{x\in\Omega}p_{x}\log\left(  \frac{p_{x}}{q_{x}%
}\right)  .
\]

\textbf{(a) The linear case}

Let $p\left(  t\right)  $ and $q\left(  t\right)  $ be two different solutions
to the ordinary differential equation%
\[
\frac{\text{d}}{\text{d}t}p\left(  t\right)  =p\left(  t\right)
\Lambda,\text{ \ }p\left(  0\right)  =q,
\]
where $\Lambda$ is the infinitesimal generator of an irreducible
continuous-time Markov process. In this case, Dupuis and Fischer
\cite{Dup:2011} indicated that%
\begin{equation}
\frac{\text{d}}{\text{d}t}R\left(  p\left(  t\right)  ||q\left(  t\right)
\right)  =-\sum_{\substack{x,y\in\Omega\\x\neq y}}\Psi\left(  \frac{p_{y}%
\left(  t\right)  q_{x}\left(  t\right)  }{p_{x}\left(  t\right)  q_{y}\left(
t\right)  }\right)  p_{x}\left(  t\right)  \frac{q_{y}\left(  t\right)
}{q_{x}\left(  t\right)  }\Lambda_{y,x}\leq0, \label{LyaF-2}%
\end{equation}
where $\Psi\left(  z\right)  =z\log z-z+1$. At the same time, they indicated
that $\frac{\text{d}}{\text{d}t}R\left(  p\left(  t\right)  ||q\left(
t\right)  \right)  =0$ if and only if $p\left(  t\right)  =q\left(  t\right)
$ for $t\geq0$; and also $\frac{\text{d}}{\text{d}t}R\left(  p\left(
t\right)  ||\pi\right)  =0$ if and only if $p\left(  t\right)  =\pi$ for
$t\geq0$.

\textbf{(b) The nonlinear case}

For the ordinary differential equation: $\frac{\text{d}}{\text{d}t}p\left(
t\right)  =p\left(  t\right)  \Gamma\left(  p\left(  t\right)  \right)  $ with
$p\left(  0\right)  =q$, Dupuis and Fischer \cite{Dup:2011} demonstrated that
the the relative entropy relation (\ref{LyaF-2}) can not be applied directly.
In this case, they first defined $\mathbf{P}^{\left(  N\right)  }\left(
t\right)  $ as the state probability of the system of $N$ big networks at time
$t\geq0$, and let $\mathbf{P}^{\left(  N\right)  }\left(  0\right)
=\otimes^{N}q$. Based on this, they gave an approximate method to construct
the Lyapunov function as follows:%
\[
F\left(  q\right)  =\lim_{N\rightarrow\infty}\lim_{T\rightarrow+\infty
}\frac{1}{N}R\left(  \mathbf{P}^{\left(  N\right)  }\left(  0\right)
||\mathbf{P}^{\left(  N\right)  }\left(  T\right)  \right)  =\lim
_{N\rightarrow\infty}\frac{1}{N}R\left(  \otimes^{N}q||\otimes^{N}\pi\right)
.
\]
For applying the relative entropy to construct a Lyapunov function, readers
may also refer to Budhiraja et al. \cite{Bud:2014, Bud:2014a} for more details.

Now, we introduce the locally stable fixed points in the set $\mathbb{S}$, and
explain convergence of the sequence $\left\{  \mu^{(N)}(t)\right\}  $ of
Markov processes.

If there exists only an element $\pi$ in the set $\mathbb{S}$, then%
\begin{equation}
\mu^{(N)}(t)\Rightarrow p\left(  t\right)  \text{, \ as }N\rightarrow\infty,
\label{FixedP-3}%
\end{equation}
and%
\begin{equation}
p\left(  t\right)  \rightarrow\pi\text{, \ as }t\rightarrow+\infty.
\label{FixedP-4}%
\end{equation}

When there exist multiple elements in the set $\mathbb{S}$, we hope to find a
similar property to that in (\ref{FixedP-3}) and (\ref{FixedP-4}). To this
end, it is necessary to introduce the locally stable fixed points, which are
first defined in Budhiraja et al. \cite{Bud:2014}. For convenience of readers,
we restate the definition of locally stable fixed points here.

We define the relative interior of $\Omega^{\Diamond}$ of $\Omega$ as%
\[
\Omega^{\Diamond}=\left\{  \mathbf{g\in\Omega}:\mathbf{g}>0\right\}  .
\]
Notice that the Markov chain $\Gamma\left(  \pi\right)  $ is irreducible, it
is easy to see from the system of fixed point equations $\pi\Gamma\left(
\pi\right)  =0$ and $\pi e=1$ that any fixed point $\pi>0$, hence we obtain
$\mathbb{S\subset}$ $\Omega^{\Diamond}$.

\begin{Def}
\label{Def:LocS}A fixed point $\pi\mathbf{\in}\Omega^{\Diamond}$ is said to be
locally stable if there exists a relative open set $\mathfrak{F}$ of $\Omega$
that contains $\pi$ and has the property that whenever $p\left(  0\right)
=q\mathbf{\in}\mathfrak{F}$, the solution $p\left(  t\right)  $ to the dynamic
system of mean-field equations (\ref{dyn-1}) and (\ref{dyn-2}) converges to
$\pi$ as $t\rightarrow+\infty$.
\end{Def}

Based on Definition \ref{Def:LocS}, every element in the set $\mathbb{S}$ is a
locally stable fixed point. In this case, let $\mathfrak{F}^{\left\langle
k\right\rangle }$ be a relative open set of $\Omega$ for $k\geq1$, and we
write%
\[
\mathbb{S=}\left\{  \pi^{\left\langle 1\right\rangle },\pi^{\left\langle
2\right\rangle },\pi^{\left\langle 3\right\rangle },\ldots\right\}  ,\text{
\ }\pi^{\left\langle k\right\rangle }\in\mathfrak{F}^{\left\langle
k\right\rangle }\text{ for }k\geq1,
\]
and%
\[
\mathbb{Q}=\left\{  q^{\left\langle 1\right\rangle },q^{\left\langle
2\right\rangle },q^{\left\langle 3\right\rangle },\ldots\right\}  ,\text{
\ }q^{\left\langle k\right\rangle }\in\mathfrak{F}^{\left\langle
k\right\rangle }\text{ for }k\geq1.
\]

If for each $k\geq1$,%
\[
\mu^{(N)}(0)\Rightarrow q^{\left\langle k\right\rangle }\in\mathfrak{F}%
^{\left\langle k\right\rangle }\text{, \ as }N\rightarrow\infty,
\]
then%
\begin{equation}
\mu^{(N)}(t)\Rightarrow p\left(  t,q^{\left\langle k\right\rangle }\right)
\text{, \ as }N\rightarrow\infty, \label{FixedP-5}%
\end{equation}
and%
\begin{equation}
p\left(  t,q^{\left\langle k\right\rangle }\right)  \rightarrow\pi
^{\left\langle k\right\rangle }\text{, \ as }t\rightarrow+\infty,
\label{FixedP-6}%
\end{equation}
where $p\left(  t,q^{\left\langle k\right\rangle }\right)  $ denotes that this
solution $p\left(  t\right)  $ depends on the initial vector
\ $q^{\left\langle k\right\rangle }\in\mathfrak{F}^{\left\langle
k\right\rangle }$, where $p\left(  0,q^{\left\langle k\right\rangle }\right)
=q^{\left\langle k\right\rangle }$.

From the above analysis, for each locally stable fixed point in $\mathbb{S}$,
we find the similar property (\ref{FixedP-5}) and (\ref{FixedP-6}) to that in
(\ref{FixedP-3}) and (\ref{FixedP-4}). Based on this, computation of the
locally stable fixed points is on a unified line, but the the initial vector
\ $q^{\left\langle k\right\rangle }\in\mathfrak{F}^{\left\langle
k\right\rangle }$ with $\mu^{(N)}(0)\Rightarrow q^{\left\langle k\right\rangle
}$ as $N\rightarrow\infty$ has a large impact on the limit $\pi^{\left\langle
k\right\rangle }$ of the solution $p\left(  t\right)  $ as $t\rightarrow
+\infty$, thus a basic task for computing the locally stable fixed points is
to find a suitable collection of the relative open sets of $\Omega$ as
follows:%
\[
\mathcal{F}=\left\{  \mathfrak{F}^{\left\langle 1\right\rangle }%
,\mathfrak{F}^{\left\langle 2\right\rangle },\mathfrak{F}^{\left\langle
3\right\rangle },\ldots\right\}  ,
\]
which sufficiently corresponds to the set $\mathbb{S=}\left\{  \pi
^{\left\langle 1\right\rangle },\pi^{\left\langle 2\right\rangle }%
,\pi^{\left\langle 3\right\rangle },\ldots\right\}  $.

In the remainder of this section, we provide several interesting open problems
with detailed interpretation.

\textbf{Open problem one: }The mean drift condition.

We consider an irreducible QBD process whose infinitesimal generator is given
by%
\[
\Gamma\left(  p\right)  =\left(
\begin{array}
[c]{ccccc}%
B_{1}\left(  p\right)  & B_{0}\left(  p\right)  &  &  & \\
B_{2}\left(  p\right)  & A_{1}\left(  p\right)  & A_{0}\left(  p\right)  &  &
\\
& A_{2}\left(  p\right)  & A_{1}\left(  p\right)  & A_{0}\left(  p\right)  &
\\
&  & \ddots & \ddots & \ddots
\end{array}
\right)  ,
\]
where $\Gamma\left(  p\right)  e=0$, the sizes of the matrices $B_{1}\left(
p\right)  $ and $A_{1}\left(  p\right)  $ are $m_{0}$ and $m$, respectively,
and the sizes of other matrices can be determined accordingly. We assume that
for any $p\in\mathfrak{P}\left(  \Omega\right)  $, the Markov process:
$A\left(  p\right)  =A_{0}\left(  p\right)  +A_{1}\left(  p\right)
+A_{2}\left(  p\right)  $, is irreducible, aperiodic and positive recurrent.
Let $\theta_{p} $ be the stationary probability vector of the the Markov
process $A\left(  p\right)  $. Then it is clear that for for any
$p\in\mathfrak{P}\left(  \Omega\right)  $, the Markov process $\Gamma\left(
p\right)  $ is positive recurrent if and only if $\theta_{p}A_{2}\left(
p\right)  e>\theta_{p}A_{0}\left(  p\right)  e$.

It is interesting to study how the mean drift condition: $\theta_{p}%
A_{2}\left(  p\right)  e>\theta_{p}A_{0}\left(  p\right)  e$ for any
$p\in\mathfrak{P}\left(  \Omega\right)  $, can influence stability or
metastability of the ordinary differential equation: $\frac{\text{d}}%
{\text{d}t}p\left(  t\right)  =p\left(  t\right)  \Gamma\left(  p\left(
t\right)  \right)  $ with $p\left(  0\right)  =q$.

\textbf{Open problem two: }The censoring Markov processes.

For the infinitesimal generator $\Gamma\left(  p\right)  $ given in
(\ref{gen-3}), it is easy to give the infinitesimal generator $\Psi_{0}\left(
p\right)  $ of the censoring Markov processes to level $0$. It is very
interesting (but difficult) to set up some useful relations of stability or
metastability between two ordinary differential equations: $\frac{\text{d}%
}{\text{d}t}p\left(  t\right)  =p\left(  t\right)  \Gamma\left(  p\left(
t\right)  \right)  $ and $\frac{\text{d}}{\text{d}t}p_{0}\left(  t\right)
=p_{0}\left(  t\right)  \Psi_{0}\left(  p\left(  t\right)  \right)  $, where
$p\left(  t\right)  =\left(  p_{0}\left(  t\right)  ,p_{1}\left(  t\right)
,p_{2}\left(  t\right)  ,\ldots\right)  $.

\section{Concluding Remarks}

This paper sets up a broad class of nonlinear continuous-time block-structured
Markov processes by means of applying the mean-field theory to the study of
big networks, and proposes some effective algorithms for computing the fixed
points of the nonlinear Markov process by means of the UL-type $RG$%
-factorization. Furthermore, this paper considers stability or metastability
of the big networks, and gives several interesting open problems with detailed
interpretation. Along such a line, there are a number of interesting
directions for potential future research, for example:

\begin{itemize}
\item providing algorithms for computing the fixed points of big networks with
multiple stable points;

\item studying the influence of the censoring Markov processes on the metastability;

\item discussing how to apply the $RG$-factorizations given in Li
\cite{Li:2010} to compute the expected residence times in the study of
metastability; and

\item analyzing some big networks with a heterogeneous geographical
environment, and set up their simultaneous systems of nonlinear Markov processes.
\end{itemize}

\section*{Acknowledgement}

We would like to thank Professor Jeffrey J. Hunter for his useful helps and
suggestions, and acknowledge his pioneering research on Markov processes,
Markov renewal processes and generalized inverses in which those interesting
results play a foundational role in the area of applied probability, such as,
the Poisson equations, the mixing times, and many stationary computation. At
the same time, this work is partly supported by the National Natural Science
Foundation of China under grant (\#71271187, \#71471160), and the Fostering
Plan of Innovation Team and Leading Talent in Hebei Universities under grant
(\# LJRC027).

\end{document}